# Surface-enhanced Raman scattering of graphene caused by self-induced nanogating by GaN nanowire array

J. Kierdaszuk,[*, ||] P. Kaźmierczak,[||] R. Bożek,[||] J. Grzonka,[i, l] A. Krajewska,[i, ¶] Z. R. Zytkiewicz,[J] M. Sobanska,[J], K. Klosek,[J] A. Wołoś,[||] M. Kamińska,[||] A. Wysmołek,[||] and A. Drabińska[||]

[||]Faculty of Physics, University of Warsaw, Pasteura 5 st., 02-093 Warsaw, Poland,

[i]Institute of Electronic Materials Technology, Wólczyńska 133 st., 01-919 Warsaw, Poland,

[l]Faculty of Materials Science and Engineering, Warsaw University of Technology, Woloska 141 st., 02-507, Warsaw, Poland.

[¶]Institute of Optoelectronics, Military University of Technology, Gen. Sylwestra Kaliskiego 2 st., 01-476, Warsaw, Poland,

[J]Institute of Physics, Polish Academy of Sciences, Al. Lotników 32/46 st., 02-668, Warsaw, Poland,

*Corresponding author. Tel. +48 573931438, E-mail: jakub.kierdaszuk@fuw.edu.pl

ABSTRACT

A constant height of gallium nitride (GaN) nanowires with graphene deposited on them is shown to have a strong enhancement of Raman scattering, whilst variable height nanowires fail to give such an enhancement. Scanning electron microscopy reveals a smooth graphene surface which is present when the GaN nanowires are uniform, whereas graphene on nanowires with substantial height differences is observed to be pierced and stretched by the uppermost nanowires. The energy shifts of the characteristic Raman bands confirms that these differences in the nanowire height has a significant impact on the local graphene strain and the carrier concentration. The images obtained by Kelvin probe force microscopy show clearly that the carrier concentration in graphene is modulated by the nanowire substrate and dependent on the nanowire density. Therefore, the observed surface enhanced Raman scattering for graphene deposited on GaN nanowires of comparable height is triggered by self-induced nano-gating to the graphene. However, no clear correlation of the enhancement with the strain or the carrier concentration of graphene was discovered.

1. INTRODUCTION

Hybrid systems containing graphene and other nanomaterials have gained a lot of attention in the recent years due to the possibility of novel applications and the potential for new physics, especially occurring at the interface between the two media. Due to its unique transport and mechanical properties graphene itself is also considered as a promising material in a number of fields.[1,2] High carrier mobility (up to 75 000 $cm^2V^{-1}s^{-1}$), low resistivity ($10^{-6}$ $\Omega$cm), large elasticity (enabling to withstand up to 25% elastic deformation) make graphene an ideal candidate to be used in flexible nanosensors.[3] Our previous studies completed on graphene deposited on gallium nitride nanowires (GaN NWs) with similar height have shown that the Raman spectra of graphene are enhanced by more than an order of magnitude, when compared to graphene deposited on a GaN epilayer.[4] This enhancement has been interpreted to be caused by Surface Enhanced Raman Scattering (SERS). There are typically two mechanisms of SERS, an electromagnetic mechanism (EM) and a chemical mechanism (CM), and it is still under debate which of them is dominant in graphene.[5] The EM is due to the enhancement of an electric field as an effect of localized surface plasmon excitation, which can lead to an increase of Raman spectrum intensity by as much as $10^{15}$ times.[6] However, that high order of enhancement is achieved only when the laser wavelength is in resonance with plasmon frequency and for Raman bands close in energy to the laser wavelength. When such resonance conditions are not fully satisfied, the observed enhancement is strongly reduced.[7] The CM theory explains the Raman spectra enhancement by chemical bonding at the surface and a charge-transfer mechanism. In this case the predicted enhancement is typically 10-100 times. The comprehensive review of the theories describing SERS mechanism can be found for example in ref. [8] Typically, the SERS effect in graphene is observed after deposition of metal particles with a well-defined size and shape and is described by the EM.[9] Although gallium nitride is a wide bandgap semiconductor, in the wurtzite structure polarization charges at high concentration are present on its surface due to the spontaneous and piezoelectric polarizations along the *c*-axis, and they are usually screened by free carriers.[10,11] Thus, vertically aligned GaN NWs can be used as a source of discontinuous self-induced nano-gates and can give rise to enhanced local electromagnetic fields via the localized surface plasmon oscillating perpendicularly and coupled to graphene surface. This vertical NW assembly provides direct access to the whole of the graphene's surface, which in turn is a big advantage when applied to a sensor structure. Studies of such structure have to take into account that NWs in contact with the graphene modify the spatial distribution of its carrier concentration and, possibly, introduce mechanical strain to the system. It has been shown that the presence of lattice distortions leads to a Dirac cone shift and overall modification of graphene's electronic properties.[12] Studies of graphene deposited on silicon pillars with equal height have revealed the pillar density impact on the strain in the graphene.[13] Graphene lying on pillars below a critical pillar pitch stay suspended and weakly interact with the substrate. Modifying the pillar density and pitch increases the strain in graphene, as detected by Raman spectroscopy mapping, leading to graphene ripples being visible in SEM images. Thus, a GaN NW array establish a promising substrate for graphene deposition, with the possibility of tailoring the graphene carrier concentration and strain, which are crucial parameters of future novel nanodevices; in particular, SERS exhibiting substrates can boost the number of possible applications in the nanosensor field.

To understand the processes responsible for the enhancement of Raman spectra in graphene, we have studied graphene deposited on NWs with a varying height distribution and different density. The topography of the samples was visualized by Scanning Electron Microscopy (SEM) and Atomic Force Microscopy (AFM). Reflectivity measurements were applied to study the light interference on the nanowire array. Finally, we used Raman spectroscopy as a versatile tool for validating graphene's structure in these assemblies. The most intense graphene 2D band of Raman spectrum is sensitive to strain, while less intensive G band depends on carrier concentration as well as strain.[14–17] Both defect bands (D and D') quantify the type and concentration of defects.[18,19] Analyzing these bands by two dimensional Raman micromapping allowed, with resolution of 100s nanometers, to trace how graphene's strain and carrier concentration are modified by its interaction with NWs. This analysis can yield information about the strain behavior and its possible correlations with carrier concentration and nanowire pattern. Distribution of carriers was also obtained with higher resolution by Kelvin probe force microscopy (KPFM). In this technique the measured electric potential between AFM needle and the sample is proportional to the Fermi level position, and consequently to the electron concentration.[20] All the performed studies enabled the relationship between nanowire substrate on graphene properties to be determined, with particular focus on the enhancement of Raman spectra.

## 2. EXPERIMENTAL DETAILS

Four samples of graphene deposited on GaN NWs were studied. GaN NWs were grown by Plasma Assisted Molecular Beam Epitaxy (PAMBE) under N-rich conditions.[21] The growth temperature and growth time were varied in order to prepare NWs with differing height distribution and different array density.[22] The diameter of NWs in all samples was approximately 40 nm. In two of the samples, the GaN NWs had near equal height. As estimated from SEM, the average height was about 900 nm and the nanowire density was 140 NWs/$\mu$m$^2$. Individual nanowires were not evenly distributed on the substrate, but they form small clusters with average distances between the clusters of approximately 250 nm. In the third sample, the nanowire density was approximately 400 NWs/$\mu$m$^2$, which is three times higher than in the previous two samples. For this sample NWs also gathered in small groups with distance between them varying from 50 to 100 nm. In this sample, the height of NWs changed smoothly from 300 to 400 nm, so the magnitude of the height variation was of approximately 100 nm. In the fourth sample, NWs with two different heights, alternating by 500 nm, were found: most of the nanowires were of approximately 1 $\mu$m height, and approximately 20% of NWs were of about 1.5 µm. The density of NWs in this sample reached 120 NWs/µm$^2$ and distances between the highest NWs were 0.5 to 1 µm.

Graphene was grown by chemical vapor deposition on copper foil, with methane as the precursor. Due to low adhesive forces between graphene and NWs, it was not possible to apply a common method of graphene transfer using a poly(methyl methacrylate) (PMMA) polymer.[23] In order to solve this problem we adopted a non-standard transfer method, with the use of a stabilizing frame.[24] For the sample with NWs of equal height (denoted as M0), we employed a marker frame, while for the other three samples (with 0, 100 and 500 nm variations in height) more stable but stiffer polymer frame made from PDMS (polydimethylsiloxane) was used. These samples were denoted as P0, P100, and P500, respectively.

The SEM images were obtained using SU8230 Hitachi microscope equipped with an in-lens secondary electron detector at 5 kV electron beam voltage. In order to visualize graphene on NWs, measurements were performed at 70° tilt of the sample. A T64000 Horiba Jobin-Yvon spectrometer equipped with an Nd:YAG laser excitation source operating at 532 nm wavelength (2.33 eV energy) was used for micro-Raman studies. Excitation power was 3 mW. Spectral resolution changed from 1.98 cm$^{-1}$ for 1200 cm$^{-1}$ Raman shift to 1.61 cm$^{-1}$ for 2800 cm$^{-1}$ Raman shift. A spatial resolution of approximately 300 nm was achieved by applying a confocal pinhole and an objective with a 100× magnification and 0.9 NA. This resolution was too low to separate images of individual NWs. On average 10 NWs were stimulated by the laser spot in the M0 and P0 samples. Due to the higher density of NWs in P100 sample, on average 28 single NWs were located in a single laser spot. In the case of P500 sample it was approximately 8 NWs under the laser spot, however only a single NW had a height of 1.5 µm. Raman micromapping was performed on area of a several square micrometers for each sample, with the 100 nm step of piezoelectric motors. In order to compare intensity of Raman scattering between individual samples, spectra were normalized with respect to the intensity of main Raman line of silicon reference sample. For reflectivity micromapping we used a halogen lamp and the same spectrometer as for Raman micromapping. The measurements were performed for wavelengths in the range of 360 to 800 nm, on 1 $\mu m^2$ sample area, and with 100 nm step between each data point. In this case an objective with a 100× magnification was also applied. AFM and KPFM measurements were carried out using Digital Instruments Multimode Atomic Force Microscope with a needle of 50 nm diameter, capable of measuring the electric potential of the specimen. The needle diameter was comparable to the nanowire diameter in all investigated samples. In KPFM measurements, due to inability to determine the reference potential, values of the measured potential are arbitrary, however their values and local variations can be compared between the investigated samples.

3. RESULTS AND DISCUSSION

Figure 1 shows the topography of the studied samples as obtained by SEM. No substantial differences between graphene on NWs with equal height transferred using two different frames (samples M0 and P0) were observed. Graphene was generally observed as plain and smooth and it lied on the apexes of all the nanowires (figure 1a, b). The small ripples were caused by slight variations in height between adjacent NWs. Larger cracks are most probably created by copper etching during the graphene transfer procedure. For the P100 sample, larger ripples caused by moderate height differences in nanowires (figure 1c) were observed. The NW density in this sample was higher than in the previous samples, but strain stretched graphene. Graphene was supported by every single nanowire including those lower in height.

Consequently, the local graphene expansion was varied. For P500 sample, presented in figure 1d, different behavior was observed. Graphene was located on the apexes of the highest NWs, with no contact to the lowest NWs. The density of NWs which were in contact with graphene is therefore much smaller than for the other three samples, and the graphene was highly expanded. For both P100 and P500 samples, graphene was pierced by some of the highest NWs. Such an effect was not observed for the M0 and P0 samples.

In the obtained Raman spectra significant variations of energy and intensity of individual Raman bands across the mapping areas were observed. Sample Raman spectra of graphene on each NWs substrate are shown in Figure 2a, while Table 1 presents the average intensity and its standard deviation of the individual Raman bands for the respective samples. The spectra of graphene on GaN NWs with equal height (M0 and P0 samples) were over an order of magnitude more intense than those coming from the two unequal height samples (P100 and P500). The spectra of P100 and P500 samples were of similar intensity, typical for graphene deposited on an epitaxial GaN layer.[4] The discussed enhancement for M0 and P0 samples was observed for the G and 2D bands, as well as for the defect D and D' bands of graphene. The detailed statistical analysis of Raman bands intensities are included in the supplementary material. Analysis of the Raman spectra showed that differing variations in height of the NWs subdivides our samples into two categories with substantially differing spectrum intensity. The first one consists of the M0 and P0 samples and the second one the P100 and P500 samples. The intensity ratios presented in Table 2 for samples within the same category (M0/P0 and P100/P500) are similar and approximately 1, while for samples from different categories were of the order of a few dozen. In this work we use notion of enhancement factor (EF) (presented in Table 2) which is defined as the ratio of Raman peak intensity of P0 (or M0) sample to the corresponding peak of P100 (or P500) sample. The highest EF was found for the 2D band (~30-40), whilst for the G band it was~15-25, and for D' band an EF of ~20-30 was observed. Interestingly, EF of the defect D band varied by a factor of 20 when compared with the P100 sample to 50 when compared with the P500 sample. We relate the difference in EF of D band to different density of NWs in contact with graphene, which for P100 sample is much larger, and for the P500 sample is smaller when compared to M0 and P0 samples.

One mechanism of the Raman scattering enhancement that could be considered is an interference of the light reflected from graphene/NWs and NWs/silicon interfaces. It has already been reported that Raman spectra of graphene deposited on a silicon substrate covered with a thin layer of silicon dioxide can be enhanced by more than an order of magnitude.[25] This has been explained as a result of interference between the light reflected from the silicon and from the silicon dioxide surface. To exclude that possibility in our samples, reflectivity micromapping measurements were performed on all samples within a surface area of 1 $\mu m^2$. As shown in figure 2b, interferences are visible for all the samples except the P500 case, i.e. graphene on NWs with the largest differences in height. Thus, as expected, high variations in NWs height suppressed the interference. Taking into account the effective refractive index of the GaN NWs layer (which is lower when compared to the planar GaN layer and is dependent on the filling factor – the ratio of NWs volume to the volume of the planar layer of the same height) it is possible to estimate height of the NWs.[26] The calculated height for M0, P0 and P100 agreed with the height of the NWs obtained from SEM measurements. It is worth to noting, that for each particular sample (for which the interferences were observed) the energy position of interference maxima did not shift when mapping on the investigated surfaces, however small variations of the average reflectivity were observed. An average reflectivity value for M0 and P0 sample was twice higher than for the P100 and P500 samples. This difference is most probably caused by higher scattering on the rougher surface. However, the magnitude of the differences in reflectivity value between the investigated samples was significantly lower than the observed EFs (table 2). Secondly, since the wavelength of laser line corresponds to the interference minimum (figure 3), the probable reason for the enhancement of the whole Raman spectrum, which is the positive interference of laser line, can be excluded. Moreover, the greatest enhanced 2D band for M0 and P0 samples was in-between interference minimum and maximum, while the positions of other three bands corresponded to the interference maximum. Therefore, we can conclude that in this case, the interference effect did not play any major role in the observed enhancement of Raman scattering for graphene on NWs with equal height.

The problem of how strain and carrier concentration in graphene influence the enhancement of Raman scattering was studied by the analysis of the dependence of G band energy on the 2D band energy (figure 3), in comparison with respective data for unstrained and undoped graphene recorded in literature studies.[14] In general, the strain changes the bond lengths in graphene and consequently impacts its phonon frequency. That effect has been clearly observed for graphene's main 2D and G bands.[14] Additionally, due to the presence of a Kohn anomaly near the $\Gamma$, K and K' points in the first Brillouin zone, phonon band energies are sensitive to the carrier concentration.[16] Experimental results obtained for gated graphene confirmed that G band energy is dependent on the carrier concentration, however the 2D band energy stays constant for doping levels lower than $5 \cdot 10^{12}$ cm$^{-2}$ for holes, and $2 \cdot 10^{13}$ cm$^{-2}$ for electrons.[16] The dependence of G band energy shift, $\Delta E_G$, on strain, $\Delta\varepsilon$, and carrier concentration, $n$, is described by the formula: [27]

$$\Delta E_G = -2\gamma_G E_G^0 \Delta\varepsilon + na \qquad (1)$$

where $\gamma_G$ is Grüneisen parameter, $a$ is a constant (equal to $7.38 \cdot 10^{13}$ cm when the band energy is given in cm$^{-1}$ units and concentration in cm$^{-2}$), and $E_G^0$ is a value of G band energy for unstrained and undoped graphene and it is equal to 1583.5 cm$^{-1}$ for 2.33 eV laser excitation energy.[14] The dependence of 2D band energy shift, $\Delta E_{2D}$, on strain is described by the formula:[27]

$$\Delta E_{2D} = -2\gamma_{2D}E_{2D}^0\Delta\varepsilon \qquad (2)$$

where $E_{2D}^0$ is a value of 2D band energy for unstrained and undoped graphene and it is equal to 2677.6 cm$^{-1}$ for 2.33 eV laser excitation energy.[14] The Grüneisen parameter describes the dependence of the phonon frequency on the crystal volume. The value of the Grüneisen parameter for G and 2D bands is strongly determined by the strain. Consequently, the absolute number of the carrier concentration can be precisely determined by formula 1 only for unstrained graphene. However, for small graphene areas, where strain is nearly constant, the changes of the G band energy are proportional to the change of the carrier concentration.

As observed in figure 3 which shows the dependence of G band energy on the 2D band energy, the data from Raman micromaps for different samples are clearly separated into different regions. We can therefore compare strain and carrier concentration of the studied samples. The strain value can be analyzed considering values of the 2D band energy (formula 2). The M0 sample is clearly the least strained from all the samples. The average compressive strain for M0 sample is ~0.03% with standard deviation ($\sigma$) of 0.02%. This result correlates well with SEM image of M0 sample showing the graphene lying smoothly on the NWs with uniform height (figure 1a). Interestingly, the average strain in P0 sample, ~0.07% ($\sigma = 0.01\%$), is more than twice higher than in the M0 sample and it has a tensile character, as opposed to the strain seen in the other samples. This can be explained by graphene stretching due to water surface tension during sample processing. The polymer frame used in case of P0 sample was not as elastic as marker frame applied for the M0 preparation, therefore it could block relaxation of graphene strain in the P0 sample. The P100 sample is more strained than M0 and less strained when compared to P500. The compressive strain for P100 and P500 samples is in average of ~0.07% with $\sigma = 0.04\%$, and 0.2% with $\sigma = 0.02\%$, respectively. In the case of the P100 sample, graphene was stretched and supported by lower as well as higher NWs. Consequently, the substrate effectively modulated the local strain and therefore the energy of the 2D band in Raman micromaps of P100 extended over a large range, indicating areas of P100 with compressive strain, but also some regions with tensile ones. The highest value of strain, observed in P500 sample, was caused by a lower density of the supporting nanowire apexes, since only the highest nanowires made contact with the graphene. From the values of the average strain in all the studied samples it is difficult to find any correlation of it with the enhancement of Raman spectra intensity, so strain does not seem to be the main factor influencing this phenomenon. Even the P0 and M0 samples, showing similar high enhancement of Raman scattering, are characterized by quite different strain. It is worth to mention that values of strain observed in our samples, agree with the values reported by other authors for graphene with small deformation transferred on nanopillars using standard PMMA method.[28] The detailed statistical analysis of Full Width at Half Maximum of Raman bands follow the trends in strain and concentration discussed before and are included in the supplementary material.

To compare the studied samples with regards to their carrier concentration we can still use the diagram presented in figure 3, but now the value of G band energy should be considered (it should be remembered that, according to formula 1, the energy of G band depends on concentration as well as on strain). To estimate the value of the carrier concentration we used the data points corresponding to small strain. Carrier concentration in M0 and P0 samples is estimated to be of the order of $10^{12}$ cm$^{-2}$.[16] The values of G band energy for P100 sample were generally the highest when compared to the rest of the samples. The estimated value of carrier concentration in P100 sample is at least an order of magnitude higher than that observed in the M0 and P0 samples. This can be related to the highest density of the NWs which support graphene and, therefore, the strongest exposition of such graphene on interactions with the NW substrate. In the case of P500 sample, its G band energy range was nearly equal to that of P0 sample, but considering the higher strain, this indicates a smaller carrier concentration when compared to the P0 sample. Therefore, we can conclude that carrier concentration in P500 is the smallest amongst all four studied samples, which is a result of the low density of NWs touching graphene. From the analysis of the 2D energy we can therefore conclude that the main factor determining the carrier concentration of graphene deposited on NWs is the density of the NWs interacting directly with graphene, with the higher density of NWs in proximate contact with graphene, the higher the carrier concentration. The carriers in graphene are therefore induced by NWs gating.

The general trends presented so far, observed by means of Raman spectroscopy, prove that the nanowire substrate and kind of transfer method strongly impacts the graphene strain and its carrier concentration. In order to understand the mechanism behind the Raman scattering enhancement, further analysis of G and 2D band energies with studies of the spatial distribution of the carrier concentration in graphene was performed. However, it must be remembered that spatial resolution of Raman micromapping was 300 nm, which is larger than the typical distance between the NWs. Generally, we can expect a spatial modulation of the carrier concentration due to the distribution of the supporting apex positions. For each sample, from the two-dimensional maps of G and 2D band energies presented in Fig. 4, we selected two areas which are depicted as diamonds or squares. The areas were selected in such a way that clear contrast in G band energy position within them could be observed. The points with relatively high G band energy, depicted by purple symbols, were adjacent to the ones with lower G band energy, which are depicted by green symbols. The local behavior of strain and the carrier concentration of these data points are also highlighted in Fig. 3. For the M0 sample, unambiguous local modulation of the carrier concentration was observed. Locally, the 2D band energy was nearly constant (meaning a similar strain), while the G band energy changed by ~0.6 cm$^{-1}$. This corresponds to variations in the carrier concentration by $10^{12}$ cm$^{-2}$.[16] We relate these variations to the distribution of the supporting apex positions, although, because of limited resolution, it is not possible to use Raman micromapping to image individual nanowires. A similar effect was also observed for the P0 sample in areas where the strain was low. For the P500 and P100 graphene on NWs with non-equal height, larger variations of both band energies were seen (positions of green symbols differed from the purple symbols). It was therefore not possible to trace the carrier concentration variations in these samples. Generally, as commented above, the local modulation of the carrier concentration in graphene can be traced by Raman spectroscopy in only the low strained samples, with an almost constant G band energy, when the changes in 2D band position can be ascribed to a variation of carrier concentration.

For further clarification of the carrier concentration's local behavior, KPFM measurements were performed (Fig. 5). This method allowed for better spatial resolution and improved observation for the correlation of carrier concentration with the positions of individual GaN nanowires. Electrical potential of a few NWs evidently not covered with graphene (as seen from topography images) was significantly higher than that of graphene (Fig. 5g). It should also be noted, that some other individual NWs observed by topography and presumably not covered with graphene, did not have clear contrast in the potential images. It seems that these nanowires were in fact covered by graphene which can tightly wrap around some of the single NWs (Fig 5g, h). The most significant observation of the KPFM analysis is however the local variations of potential values on the graphene itself. The KPFM results show that local spatial modulation of potential on the graphene surface correlates well with the topography of the samples. We clearly observe the expected modulation of the carrier concentration (seen by electric potential changes) related to the positions of the apexes in contact with the graphene layer (seen in the topography image). The spatial contrasts are more visible in figures which present the potential distribution (Fig. 5). For the P100 and P500 samples (figure 5c, d respectively), the graphene surface had an increased roughness, when compared to the M0 and P0 samples (figure 5a, b), what correlated with the SEM observations. The average value of the electric potential was highest in the case of the P100 sample (~0.23 V) comparing to 0.10 V for M0 and P0 samples, and 0.06 V for P500 sample. This suggests that the highest carrier concentration is in P100 sample and the lowest is in the P500 sample, this is in agreement with the trends in carrier concentration observed by Raman studies discussed before. This effect is explained by the highest density of nanowires being in contact with graphene in P100 sample and the lowest in P500, so the average value of potential and furthermore the carrier concentration was strongly influenced by the total area of graphene/NW interface. The local variations of the potential can be better traced in the cross-sections of electrical potential, which are presented in figures 5 i, j, k and l. The local variations of the potential, corresponding to the contact with graphene and the individual nanowire apexes, had an average amplitude 0.02 V for the M0 and P0 samples. A slightly higher amplitude was observed for the P100 sample. The spatial period of these variations changed from ~100 nm for the P100 sample to ~250 nm for the P0 and M0 samples. These values agreed with the distances between the adjacent NW groups, as observed in the SEM images (Fig. 1). In the case of the P500 sample, the number of NWs in contact with graphene was small, but still interaction between the individual NWs with the graphene could be traced in area close to the highest nanowire apexes seen in the topography image. The spatial period of these interactions was ~1 μm and corresponded to the distance between the highest apexes directly in contact with the graphene.

For the origin of the Raman scattering enhancement in graphene deposited on GaN NWs, we have to acknowledge the carrier modulation by the nanowire apexes, which is clearly visible with high spatial resolution by KPFM. This points to a SERS mechanism. However, observation of the enhancement, in only part of the samples, indicates that the NW substrate itself is not sufficient condition of the Raman scattering enhancement in graphene. The analysis of graphene deposited on NWs with differing height variations showed that strain, carrier concentration and density of NWs in contact with graphene varied between samples and seem to be important factors affecting the strength of the Raman scattering enhancement. The enhancement was observed for the samples deposited on NWs with the same height. However, these samples were not characterized either by the lowest strain or by the highest carrier concentration. For the P100 sample, for which no enhancement was observed, the highest carrier concentration from all the studied samples was observed and a similar strain value (although of opposite sign) as observed in the P0 sample, which exhibited an enhancement. It is probable that NW density may be an important factor affecting the light interaction with graphene layer, leading to the Raman scattering. However, further studies are needed to understand how the NWs density influences Raman scattering enhancement.

4. SUMMARY

We conclude that the main cause of Raman scattering enhancement observed for graphene deposited on GaN NW layer is due to SERS originating from local GaN nanogating. The performed experiments prove that modulation of the carrier concentration is correlated with positions of the NW apexes in contact with the graphene layer. Only the deposition on nanowires with equal height resulted in a strong Raman scattering enhancement. KPFM measurements showed that the density of nanowire substrate impacts on the graphene carrier concentration. Although our studies show that the nanowire substrate introduces inhomogeneous strain to the graphene, it seems not to have any particular effect on the enhancement factor. Analysis of substrates with different NW density shows their impact on the self-induced nanogating, but the eventual influence of NW density on the Raman enhancement requires further work.

Acknowledgments

This work was partially supported by the Ministry of Science and Higher Education in years 2015-2019 as a research grant "Diamond Grant" (No. DI2014 015744). GaN nanowires were grown within the Polish National Science Centre Grant No. UMO-2016/21/N/ST3/03381. The authors wish to acknowledge I. Pasternak and W. Strupinski for graphene growth.

BIBLIOGRAPHY

[1]     S. V. Morozov, K.S. Novoselov, M.I. Katsnelson, F. Schedin, D.C. Elias, J.A. Jaszczak, A.K. Geim, Giant Intrinsic Carrier Mobilities in Graphene and Its Bilayer, Phys. Rev. Lett. 100 (2007) 16602. doi:10.1103/PhysRevLett.100.016602.

[2]     C. Lee, X. Wei, J.W. Kysar, J. Hone, Measurement of the elastic properties and intrinsic strength of monolayer graphene., Science. 321 (2008) 385–388. doi:10.1126/science.1157996.

[3]     Y.-M. Chen, S.-M. He, C.-H. Huang, C.-C. Huang, W.-P. Shih, C.-L. Chu, J. Kong, J. Li, C.-Y. Su, Ultra-large suspended graphene as a highly elastic membrane for capacitive pressure sensors, Nanoscale. 8 (2016) 3555–3564. doi:10.1039/C5NR08668J.

[4]     J. Kierdaszuk, P. Kaźmierczak, A. Drabińska, K. Korona, A. Wołoś, M. Kamińska, A. Wysmołek, I. Pasternak, A. Krajewska, K. Pakuła, Z.R. Zytkiewicz, Enhanced Raman scattering and weak localization in graphene deposited on GaN nanowires, Phys. Rev. B. 92 (2015) 195403. doi:10.1103/PhysRevB.92.195403.

[5]     D. Cialla, A. März, R. Böhme, F. Theil, K. Weber, M. Schmitt, J. Popp, Surface-enhanced Raman spectroscopy (SERS): progress and trends, Anal. Bioanal. Chem. 403 (2012) 27–54. doi:10.1007/s00216-011-5631-x.

[6]     S. Nie, S. Emory, Probing Single Molecules and Single Nanoparticles by Surface-Enhanced Raman Scattering, Science. 275 (1997) 1102–6. doi:10.1126/science.275.5303.1102.

[7]     Y. Zhao, X. Liu, D.Y. Lei, Y. Chai, Effects of surface roughness of Ag thin films on surface-enhanced Raman spectroscopy of graphene: spatial nonlocality and physisorption strain, Nanoscale. 6 (2014) 1311–1317. doi:10.1039/C3NR05303B.

[8]     S.-Y. Ding, E.-M. You, Z.-Q. Tian, M. Moskovits, Electromagnetic theories of surface-enhanced Raman spectroscopy, Chem. Soc. Rev. 46 (2017) 4042–4076. doi:10.1039/C7CS00238F.

[9]     F. Schedin, E. Lidorikis, A. Lombardo, V.G. Kravets, A.K. Geim, A.N. Grigorenko, K.S. Novoselov, A.C. Ferrari, Surface-Enhanced Raman Spectroscopy of Graphene, ACS Nano. 4 (2010) 5617–5626. doi:10.1021/nn1010842.

[10]    O. Ambacher, J. Smart, J.R. Shealy, N.G. Weimann, K. Chu, M. Murphy, W.J. Schaff, L.F. Eastman, R. Dimitrov, L. Wittmer, M. Stutzmann, W. Rieger, J. Hilsenbeck, Two-dimensional electron gases induced by spontaneous and piezoelectric polarization charges in N- and Ga-face AlGaN/GaN heterostructures, J. Appl. Phys. 85 (1999) 3222. doi:10.1063/1.369664.

[11]    T. Takeuchi, C. Wetzel, S. Yamaguchi, H. Sakai, H. Amano, I. Akasaki, Y. Kaneko, S. Nakagawa, Y. Yamaoka, N. Yamada, Determination of piezoelectric fields in strained GaInN quantum wells using the quantum-confined Stark effect, Appl. Phys. Lett. 73 (1998) 1691–1693. doi:10.1063/1.122247.

[12]    M. Huang, H. Yan, T.F. Heinz, J. Hone, Probing strain-induced electronic structure change in graphene by Raman spectroscopy, Nano Lett. 10 (2010) 4074–4079. doi:10.1021/nl102123c.

[13]   A. Reserbat-Plantey, D. Kalita, L. Ferlazzo, S. Autier-Laurent, K. Komatsu, C. Li, R. Weil, Z. Han, A. Ralko, L. Marty, S. Guéron, N. Bendiab, H. Bouchiat, V. Bouchiat, Strain superlattices and macroscale suspension of Graphene induced by corrugated substrates, Nano Lett. 14 (2014) 5044–5051. doi:10.1021/nl5016552.

[14]   T.M.G. Mohiuddin, A. Lombardo, R.R. Nair, A. Bonetti, G. Savini, R. Jalil, N. Bonini, D.M. Basko, C. Galiotis, N. Marzari, K.S. Novoselov, A.K. Geim, A.C. Ferrari, Uniaxial strain in graphene by Raman spectroscopy: G peak splitting, Grüneisen parameters, and sample orientation, Phys. Rev. B - Condens. Matter Mater. Phys. 79 (2009) 205433. doi:10.1103/PhysRevB.79.205433.

[15]   J. Zabel, R.R. Nair, A. Ott, T. Georgiou, A.K. Geim, K.S. Novoselov, C. Casiraghi, Raman spectroscopy of graphene and bilayer under biaxial strain: Bubbles and balloons, Nano Lett. 12 (2012) 617–621. doi:10.1021/nl203359n.

[16]   A. Das, S. Pisana, B. Chakraborty, S. Piscanec, S.K. Saha, U. V Waghmare, K.S. Novoselov, H.R. Krishnamurthy, A.K. Geim, A.C. Ferrari, A.K. Sood, Monitoring dopants by Raman scattering in an electrochemically top-gated graphene transistor., Nat. Nanotechnol. 3 (2008) 210–215. doi:10.1038/nnano.2008.67.

[17]   K. Grodecki, R. Bozek, W. Strupinski, A. Wysmolek, R. Stepniewski, J.M. Baranowski, Micro-Raman spectroscopy of graphene grown on stepped 4H-SiC (0001) surface, Appl. Phys. Lett. 100 (2012) 261604. doi:10.1063/1.4730372.

[18]   L.G. Cançado, A. Jorio, E.H.M. Ferreira, F. Stavale, C.A. Achete, R.B. Capaz, M.V.O. Moutinho, A. Lombardo, T.S. Kulmala, A.C. Ferrari, Quantifying defects in graphene via

Raman spectroscopy at different excitation energies, Nano Lett. 11 (2011) 3190–3196. doi:10.1021/nl201432g.

[19] A. Eckmann, A. Felten, A. Mishchenko, L. Britnell, R. Krupke, K.S. Novoselov, C. Casiraghi, Probing the nature of defects in graphene by Raman spectroscopy, Nano Lett. 12 (2012) 3925–3930. doi:10.1021/nl300901a.

[20] M. Nonnenmacher, M.P. O'Boyle, H.K. Wickramasinghe, Kelvin probe force microscopy, Appl. Phys. Lett. 58 (1991) 2921–2923. doi:10.1063/1.105227.

[21] A. Wierzbicka, Z.R. Zytkiewicz, S. Kret, J. Borysiuk, P. Dluzewski, M. Sobanska, K. Klosek, A. Reszka, G. Tchutchulashvili, A. Cabaj, E. Lusakowska, Influence of substrate nitridation temperature on epitaxial alignment of GaN nanowires to Si(111) substrate., Nanotechnology. 24 (2013) 35703. doi:10.1088/0957-4484/24/3/035703.

[22] M. Sobanska, K.P. Korona, Z.R. Zytkiewicz, K. Klosek, G. Tchutchulashvili, Kinetics of self-induced nucleation and optical properties of GaN nanowires grown by plasma-assisted molecular beam epitaxy on amorphous AlxOy, J. Appl. Phys. 118 (2015) 184303. doi:10.1063/1.4935522.

[23] T. Ciuk, I. Pasternak, A. Krajewska, J. Sobieski, P. Caban, J. Szmidt, W. Strupinski, Properties of Chemical Vapor Deposition Graphene Transferred by High-Speed Electrochemical Delamination, J. Phys. Chem. C. 117 (2013) 20833–20837. doi:10.1021/jp4032139.

[24] I. Pasternak, A. Krajewska, K. Grodecki, I. Jozwik-Biala, K. Sobczak, W. Strupinski, Graphene films transfer using marker-frame method, AIP Adv. 4 (2014) 97133.

doi:10.1063/1.4896411.

[25]   X. Ling, J. Zhang, Interference phenomenon in graphene-enhanced Raman scattering, J. Phys. Chem. C. 115 (2011) 2835–2840. doi:10.1021/jp111502n.

[26]   H.-Y. Chen, H.-W. Lin, C.-Y. Wu, W.-C. Chen, J.-S. Chen, S. Gwo, Gallium nitride nanorod arrays as low-refractive-index transparent media in the entire visible spectral region., Opt. Express. 16 (2008) 8106–8116. doi:10.1364/OE.16.008106.

[27]   J.M. Urban, P. Dąbrowski, J. Binder, M. Kopciuszyński, A. Wysmołek, Z. Klusek, M. Jałochowski, W. Strupiński, J.M. Baranowski, Nitrogen doping of chemical vapor deposition grown graphene on 4H-SiC (0001), J. Appl. Phys. 115 (2014) 233504. doi:10.1063/1.4884015.

[28]   H. Mi, S. Mikael, C.-C. Liu, J.-H. Seo, G. Gui, A.L. Ma, P.F. Nealey, Z. Ma, Creating periodic local strain in monolayer graphene with nanopillars patterned by self-assembled block copolymer, Appl. Phys. Lett. 107 (2015) 143107. doi:10.1063/1.4932657.

List of figures

Table of contents

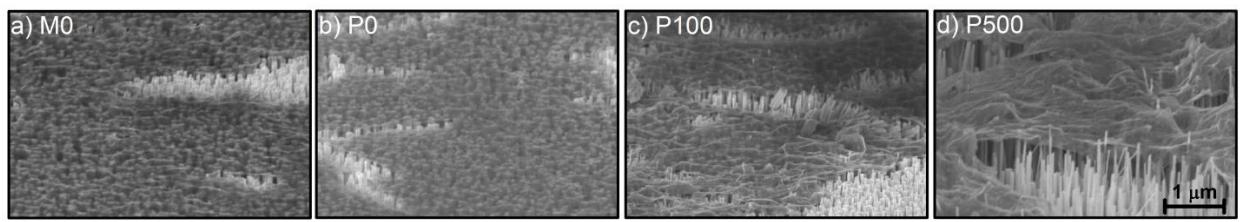

**Figure 1.** Scanning electron microscopy images of graphene on NWs with (a) 0 nm variations in height and graphene transferred using marker frame – M0, (b) 0 nm variations in height and graphene transferred using polymer frame – P0, (c) 100 nm variations in height and graphene transferred using polymer frame – P100, and (d) 500 nm variations in height and graphene transferred using polymer frame – P500.

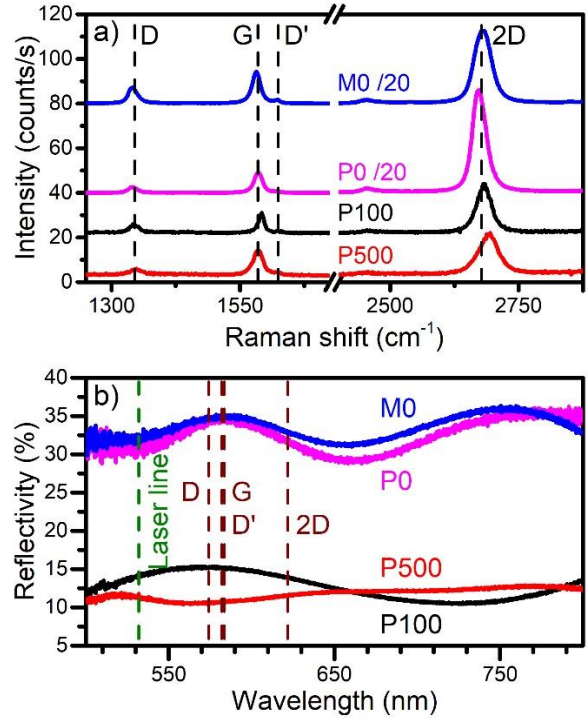

**Figure 2.** (a) Sample Raman spectra of graphene deposited on NWs with different variations in height (M0, P0, P100 and P500 samples). Due to strong enhancement for M0 and P0 samples, intensities of their spectra were divided by 20. (b) Sample reflectivity spectra of graphene on nanowires with different variations in height. The horizontal lines mark the wavelength of laser line and respective Raman bands.

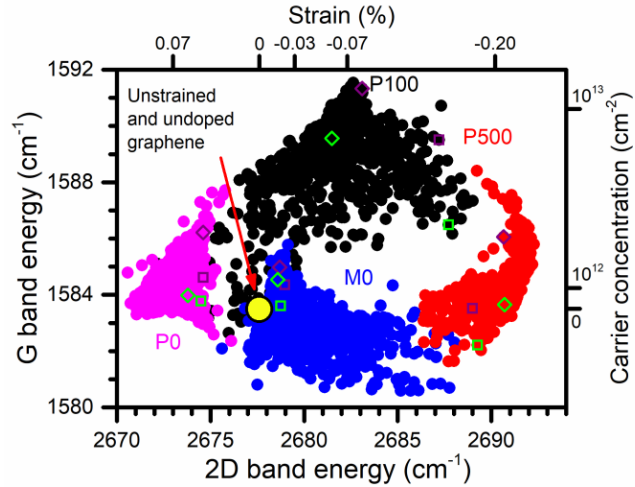

**Figure 3.** Diagram of G band energy on 2D band energy dependence for the investigated samples. Open square and diamond symbols represent data marked with the same symbols on two dimensional maps in figure 4. The yellow circle corresponds to energy values for unstrained and undoped graphene as reported in literature.[14] Sample values of strain and carrier concentration (for unstrained graphene) are presented in top and right axis respectively.

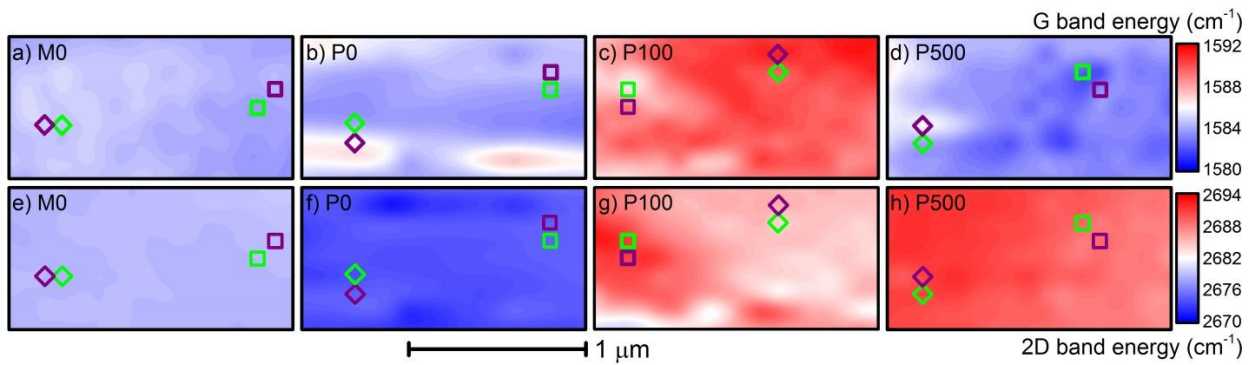

**Figure 4.** Two dimensional maps of G band energy for samples: (a) M0, (b) P0, (c) P100, (d) P500, and two dimensional maps of 2D band energy for samples: (e) M0, (f) P0, (g) P100, (h) P500. Data corresponding to areas depicted by square and diamond symbols are also presented in figure 3.

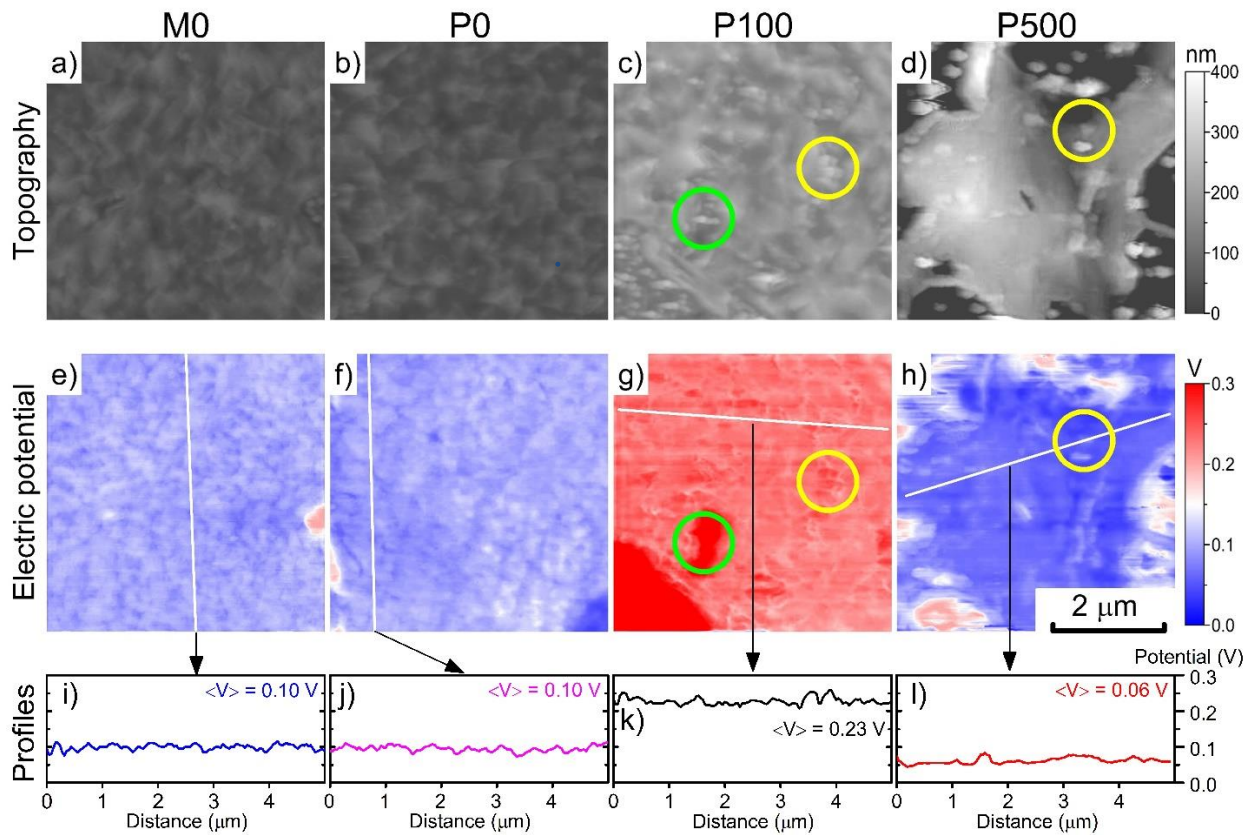

**Figure 5.** Kelvin probe force microscopy images of the topography of graphene on NWs samples: (a) M0, (b) P0, (c) P100, (d) P500, and images of potential on the surface of these samples: (e) M0, (f) P0, (g) P100, (h) P500. Area of P100 sample not covered with graphene is depicted by green circle. Area of P100 and P500 sample where graphene is tightly wrapped around single NWs is depicted by yellow circles. White lines correspond to representative cross-sections profiles for samples: (i) M0, (j) P0, (k) P100, (l) P500. ⟨V⟩ indicates average value of potential.

List of tables:

Table 1. Average intensities ($I$), in counts per second, of graphene Raman bands and their standard deviations ($\sigma$).

|  | M0 | P0 | P100 | P500 |
|---|---|---|---|---|
| $I_G$ | 218 | 156 | 9 | 11 |
| $\sigma_{I_G}$ | 93 | 29 | 5 | 2 |
| $I_{2D}$ | 564 | 792 | 21 | 19 |
| $\sigma_{I_{2D}}$ | 156 | 152 | 8 | 3 |
| $I_D$ | 101 | 87 | 4 | 2.0 |
| $\sigma_{I_D}$ | 41 | 30 | 1 | 0.3 |
| $I_{D'}$ | 16 | 11 | 0.6 | 0.6 |
| $\sigma_{I_{D'}}$ | 4 | 3 | 0.1 | 0.2 |

Table 2. Averaged relative enhancement factors of Raman bands intensity in investigated samples.

|  | M0/P100 | M0/P500 | P0/P100 | P0/P500 | M0/P0 | P100/P500 |
|---|---|---|---|---|---|---|
| G | 24 | 20 | 17 | 14 | 1.4 | 0.8 |
| 2D | 27 | 30 | 38 | 42 | 0.7 | 1.1 |
| D | 25 | 51 | 22 | 44 | 1.2 | 2 |
| D' | 27 | 27 | 18 | 18 | 1.5 | 1 |

Supplementary material

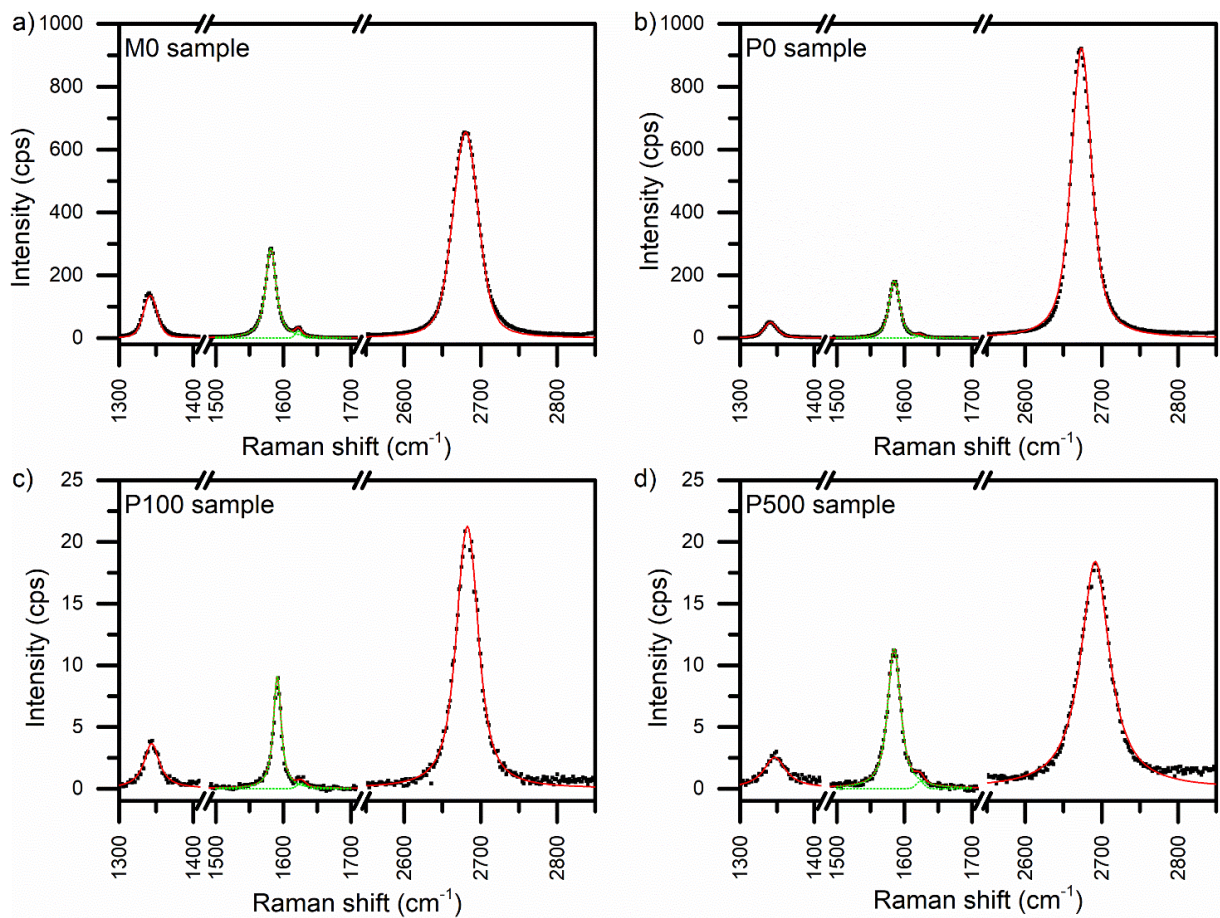

**Supplementary figure 1.** Sample Raman spectra (black dots) with appropriate fits (red curves) for samples: (a) M0, (b) P0, (c) P100, (d) P500. Green dashed lines show deconvolution of G and D' bands. All bands in all acquired spectra were fitted using Voigt profile.

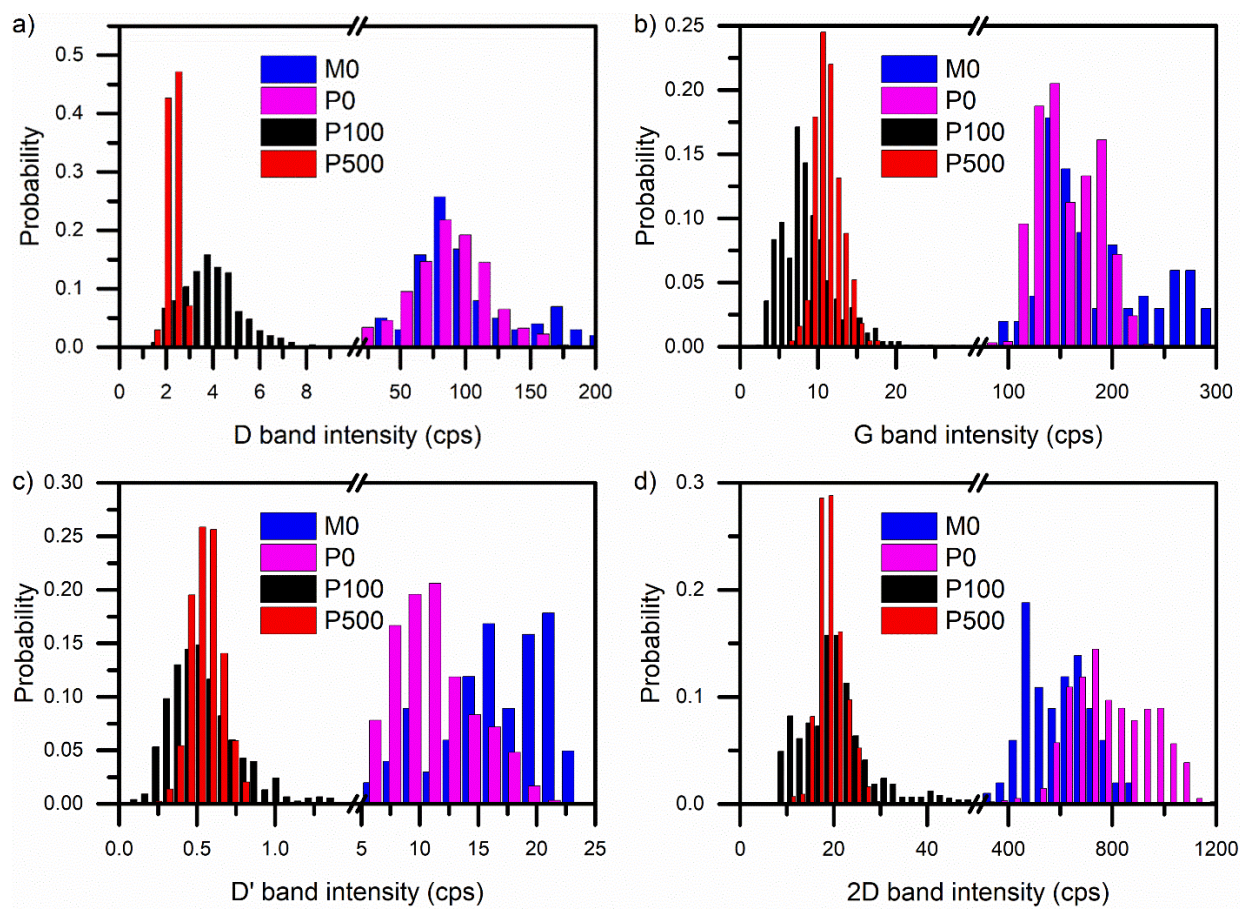

**Supplementary figure 2.** Histograms of intensity of Raman bands: (a) D, (b) G, (c) D', (d) 2D for all samples.

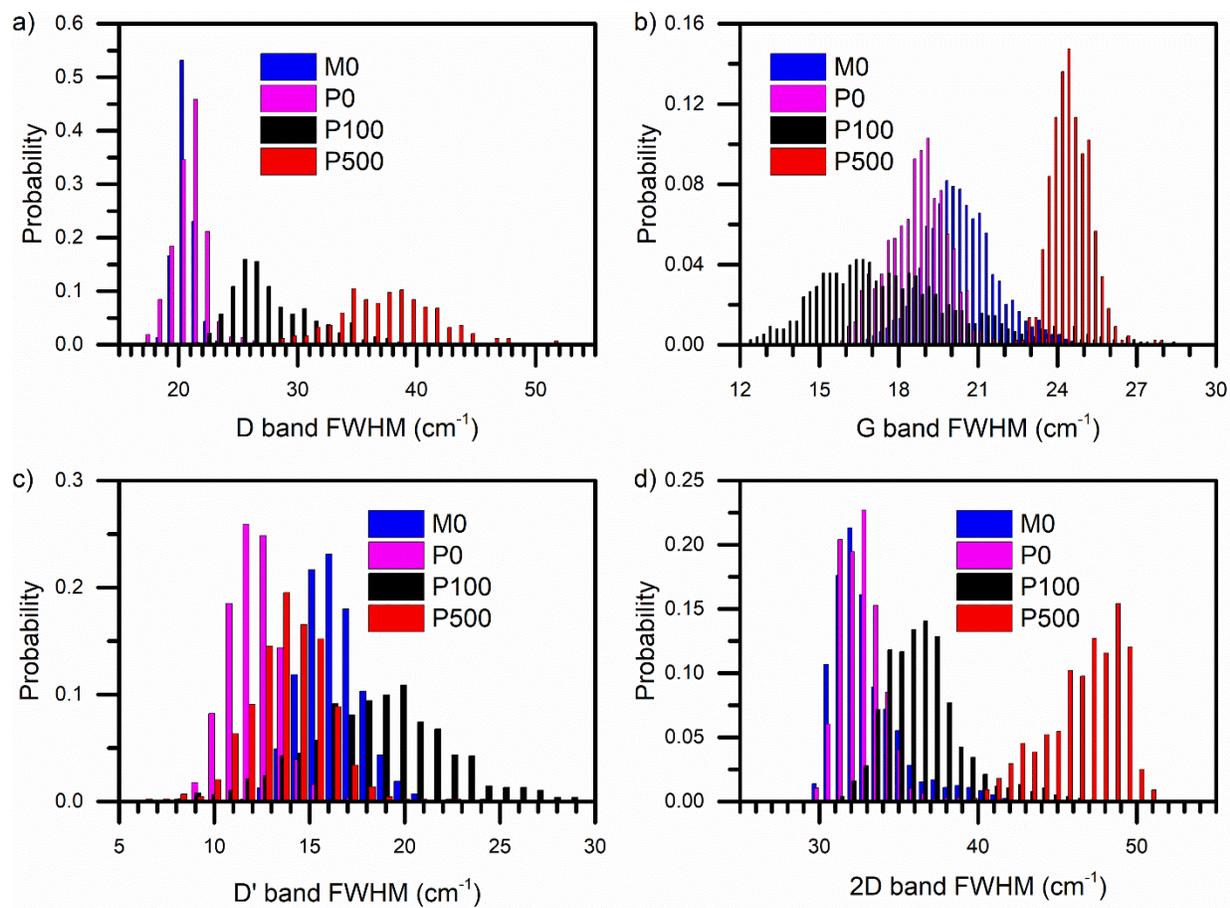

**Supplementary figure 3.** Histograms of Full Width at Half Maximum (FWHM) of Raman bands: (a) D, (b) G, (c) D', (d) 2D for all samples.